\documentclass[letterpaper, 10 pt , draftcls, onecolumn]{ieeeconf}

\IEEEoverridecommandlockouts                              

\usepackage{graphicx} 
\usepackage{epsfig} 
\usepackage{cite} 
\usepackage{amsmath} 
\usepackage{amssymb}  
\usepackage{mathrsfs}
\usepackage{color}
\usepackage{nicefrac}
\usepackage[linesnumbered, ruled, vlined]{algorithm2e}
\usepackage{mathtools}
\mathtoolsset{showonlyrefs}

\usepackage{amsthm}
\graphicspath{{.}}

\theoremstyle{plain}
\newtheorem{thm}{Theorem}[subsection]
\usepackage{etoolbox}
\usepackage{float}
\usepackage{tabularx}

\def\th@definition{
	\thm@headfont{\itshape} 
	\thm@notefont{} 
}
\makeatother

\theoremstyle{definition}
\newtheorem{dfn}[thm]{Definition}

\newtheorem{lemma}{Lemma}

\newtheorem{problem}{Problem}
\newtheorem{theorem}{Theorem}

\newtheorem{assumption}{Assumption}

\newcommand{\barsig}{\overline{\Sigma}}
\newcommand{\tell}{\tilde{\ell}}

\newcommand{\adj}{\textnormal{Adj}}

\newcommand{\tr}{\textnormal{tr}}

\makeatletter
\patchcmd{\@makecaption}
{\scshape}
{}
{}
{}
\makeatother

\title{\LARGE \bf
Error Bounds and Guidelines for Privacy Calibration in Differentially Private Kalman Filtering
}

\author{Kasra Yazdani$^{\ast}$ and Matthew Hale$^{\ast}$
\thanks{$^{\ast}$Kasra Yazdani and Matthew Hale are with the Department of  Mechanical and Aerospace Engineering at the University of Florida, Gainesville, FL USA. Emails: \texttt{\{kasra.yazdani,matthewhale\}@ufl.edu.} 
}
}

\begin{document}

\maketitle

\begin{abstract}
	Differential privacy has emerged as a formal framework
	for protecting sensitive information in control systems.
	One key feature is that it is immune to post-processing, which means that arbitrary
	post-hoc computations can be performed on privatized data without weakening differential privacy.
	It is therefore common to filter private data streams. To characterize this setup, in this paper we
	present error and entropy bounds for Kalman filtering
	differentially private
	state trajectories. We consider systems in which an output trajectory is privatized in order to protect the state trajectory that produced it. We provide bounds
	on \emph{a priori} and \emph{a posteriori} error and differential entropy of
	a Kalman filter which is processing the privatized output trajectories. 
	Using the error bounds we develop, we then provide
	guidelines to calibrate privacy levels
	in order to keep filter error within pre-specified bounds.
	Simulation results are presented to demonstrate these
	developments. 
\end{abstract}
\section{Introduction}

Emerging technologies such as smart cities~\cite{eckoff2018},  intelligent transportation systems~\cite{kargl2013}, and smart power grids~\cite{McDaniel2000} all promise to improve user services with a data-driven approach. A unifying theme in these applications is the reliance on user data in driving decisions about traffic routing, power generation, and other system behaviors. Simultaneously, these data streams have been shown to be quite revealing about users, potentially disclosing their daily habits~\cite{Lisovich2008} and their locations~\cite{zhang2017security}. Thus there has arisen a need for user privacy. Given the need for data in decision making, privacy must also preserve the usefulness of privatized data to its recipient.

In recent years, differential privacy has become a common framework for privacy of this kind. Differential privacy started in the database literature~\cite{Dwork2014algorithmic} and is used to privatize database entries when database queries are made. 
More recently, differential privacy has been extended to trajectories arising in systems and control in~\cite{LeNy2014}, where the goal is to preserve the privacy of whole trajectories
of data as they are generated.

Differential privacy is simple to implement because it merely requires adding noise to sensitive data (or functions of sensitive data). 
Differential privacy has several other properties that make it useful in control system applications. In particular, it is immune to post-processing and robust to side-information, which means that its privacy guarantees are invariant under 
post-hoc transformations and that they are not weakened by much with the availability of auxiliary information~\cite{Dwork2014algorithmic}. As a result, further computations can be performed once data is privatized without harming differential privacy's guarantees. Of course, adding noise changes the accuracy of these computations relative to their noise-free counterparts, and 
the effects of differential privacy have been investigated in several contexts, 
e.g.,~\cite{alvim2012,kifer2011,Dwork2014algorithmic,wang2018,LeNy2014,cortes2016}. 

%

Data-driven systems must share data to run. In combination with the vulnerability of dynamic data streams, this need has stimulated the use of differential privacy along with model-based state estimators to both protect sensitive data and make useful control decisions in the presence of privacy.  
Differential privacy's immunity to post-processing means that state estimation and filtering can be performed freely without threatening the privacy guarantees
of a system's data. 
The Kalman filter is a widely used state estimator which has been shown to improve the utility of privatized data in various settings~\cite{degue2017,LeNy2014}.
One common approach to differentially private Kalman filtering is to have systems add noise directly to system outputs; this approach is broadly termed ``input perturbation,'' because individual systems add noise to the inputs of the Kalman filter. This approach has the advantage of privatizing all data before it is ever shared,
eliminating the need for a trusted aggregator. 

In this paper, we are interested in privacy over long time horizons, as in smart grids and other systems that will be active for a long time. We therefore consider a Kalman filter in \emph{steady state} and we analyze the accuracy of filtering private data under the input perturbation paradigm. 
We consider a system with discrete-time dynamics, and we protect its state trajectory by adding noise to its outputs at each
point in time. 
We quantify the effects of privacy in two ways. First, we use differential entropy to quantify the information content of the privatized output trajectories. Second, we investigate the practical effects of privacy by bounding the mean squared error of an external observer's estimate of the system's states.

While the computer science literature has devised methods to calibrate privacy based on analyzing static data \cite{hsu2014}, to the best of our knowledge, no such systematic study has been undertaken in the control theory literature for trajectory-valued data. We therefore explore the relationship between privacy levels, the amount of information revealed, and the accuracy of estimates based on privatized data. We do so in terms of system properties and dynamics, thereby directly linking control theoretic information with privacy of system trajectories. 
Based on the bounds we derive, 
we provide guidelines for selecting one's privacy level based on the downstream filtering error it induces. 
Through doing so, we provide the ability to calibrate one's privacy levels based on conventional control-theoretic concerns (i.e., filtering error), thereby
enabling meaningful privacy calibration without requiring in-depth knowledge of differential privacy. 
In addition, this paper differs from \cite{LeNy2014,leny14b} because those papers design private filters, whereas we characterize a common Kalman filter setup rather than designing novel filtering strategies. 

The rest of the paper is organized as follows. Section ~\ref{sec:privacyReview_problemFormulation} provides the necessary background for differential privacy and sets up the problem statement. In Section~\ref{sec:privacyImplementation}, we outline the privacy implementation for our problem and briefly review Kalman filtering. Section~\ref{sec:results} presents the first main results of the paper, which are bounds on the differential entropy and MSE of state estimates based on  private data. In Section~\ref{sec:guidelines}, we provide guidelines for calibrating privacy levels based on pre-specified error bounds. Next, we present numerical simulations in Section~\ref{sec:casestudy}, and Section~\ref{sec:conclusions} concludes the paper.

\section{Background on Differential Privacy and Problem Formulation}\label{sec:privacyReview_problemFormulation}
In this section we first briefly review the relevant privacy background as it pertains to private Kalman filtering specifically,
and we refer the reader to~\cite{LeNy2014} for a complete exposition. Then we state the problem that is the subject of the paper. 

\subsection{Review of Differential Privacy}\label{subsec:privacy_review}
Differential privacy is a quantitative and statistical means of protecting data. Differential privacy makes it unlikely that an adversary or eavesdropper 
can make high-fidelity inferences about individuals by looking at their privatized data. It is enforced by adding noise to sensitive data (or functions thereof). Control theory provides many techniques that compensate for noise, making differential privacy a natural choice in control-theoretic settings.

Differential privacy is immune to post-processing, which means that transforming private data does not harm its privacy guarantees. In particular, filtering private trajectories does not degrade the protections of differential privacy. The guarantees of differential privacy are also robust to mechanism knowledge, which means that adversaries do not gain any advantage if they know the mechanism used to privatize data~\cite{Dwork2006_3,Dwork2014algorithmic}.

In this paper, we consider a system with discrete-time dynamics. The state trajectory of the system is sensitive, and therefore it needs to be protected. Denote the system's state
trajectory by $x$. The $k^{th}$ element of 
$x$ is denoted by $x\left(k\right)\in\mathbb{R}^{n}$
for some $n\in\mathbb{N}$. The notion of differential privacy in this paper follows the definition of differential privacy for trajectories introduced in \cite{LeNy2014}. Differential privacy can be used to ensure that an adversary is unlikely to determine either the input or state trajectory of a system, and in this paper we implement differential privacy to protect state trajectories.
  
In this work, we consider the so-called ``input perturbation'' approach to differential privacy. This means that a  system will directly add noise to its own outputs before sharing them, and this has the advantage of masking sensitive data before it is shared. Formally, a system's state trajectory will be made approximately indistinguishable from other nearby state trajectories which that system could have produced; the notions of ``nearby'' and ``approximately indistinguishable'' are formalized below in Definitions~\ref{def:adj} and~\ref{dfn:differential privacy}.

We consider vector-valued trajectories of the general form
${
Z=\left(Z\left(1\right),Z\left(2\right),\dots,Z\left(k\right),\dots\right),
}$
 where $Z\left(k\right)\in\mathbb{R}^{d}$ for all $k$. We also
use the~$\ell_p$-norm
${
\left\Vert Z\right\Vert _{\ell_{p}}:=\left(\sum_{k=1}^{\infty}\left\Vert Z\left(k\right)\right\Vert _{p}^{p}\right)^{\frac{1}{p}},
}$
where $\left\Vert .\right\Vert _{p}$ is the ordinary $p\text{-norm}$
on $\mathbb{R}^{d}$. We further define the set 
$
\ell_{p}^{d}:=\left\{ Z\mid Z\left(k\right)\in\mathbb{R}^{d},\;\left\Vert Z\right\Vert _{\ell_{p}}<\infty\right\} .
$

The state trajectory $x$ is contained
in the set $\tilde{\ell}_{2}^{n}$, which is the set of sequences
of vectors in $\mathbb{R}^{n}$ whose finite truncations are all
in $\ell_{2}^{n}$. Formally, we define the truncation operator
$P_{T}$ over trajectories according to
\begin{equation}
P_{T}\left[x\right]=\begin{cases}
x\left(k\right) & k\le T\\
0 & k>T
\end{cases}\text{,}
\end{equation}
and we say that $x\in\tilde{\ell}_{2}^{n}$ if and only if
$P_{T}[x]\in\ell_{2}^{n}$ for all $T\in\mathbb{N}$.

A differential privacy mechanism makes adjacent trajectories
produce outputs which are similar in a precise sense. The choice of adjacency relation is a key
part of any differential privacy implementation because it specifies which sensitive pieces of data must be
made approximately indistinguishable.  
To formulate differential privacy for trajectories, we next define the adjacency relation
over the space $\tilde{\ell}_{2}^{n}$
defined above. 
\begin{dfn} \label{def:adj} \emph{(Adjacency) }
Fix an adjacency parameter ${B > 0}$. 
	The adjacency relation $\adj_{B}$ is defined
	for all $v, w \in \tell^{n}_2$ as
	\begin{equation}
	\adj_{B}(v, w) = \begin{cases}
	1 & \|v - w\|_{\ell_2} \leq B \\
	0 & \text{otherwise}. 	\tag*{$\triangle$}
	\end{cases}
	\end{equation}
	
\end{dfn}

Two state trajectories of the system are thus adjacent if the~$\ell_2$ distance between them
is not more than~$B$. Differential privacy must therefore make the system's
state trajectory approximately indistinguishable from all others contained
in an~$\ell_2$-ball of radius~$B$ centered on its actual trajectory.

Next is a formal definition of differential privacy for dynamical systems which specifies the probabilistic guarantees of privacy. 
To state it, we will use a probability space $\text{(\ensuremath{\Omega},\ensuremath{\mathcal{F}},\ensuremath{\mathbb{P}}})$.
This definition considers outputs in the space $\tilde{\ell}_{2}^{q}$
and uses a $\sigma\text{-algebra}$ over $\tilde{\ell}_{2}^{q}$,
denoted $\Sigma_{2}^{q}$, construction of which can be found in~\cite{LeNy2014}.

\begin{dfn}\label{dfn:differential privacy}
	\emph{($(\epsilon,\delta)$-Differential Privacy)} 
	Let\emph{ $\epsilon>0$
		}and $\delta\in\left(0,\nicefrac{1}{2}\right)$ be given. A mechanism
		$M:\tilde{\ell}_{2}^{n}\times\Omega\rightarrow\tilde{\ell}_{2}^{q}$
		is $\left(\epsilon,\delta\right)$-differentially private if
		for all adjacent ${ x,x'\in\tilde{\ell}_{2}^{n} }$,
		we have
		\begin{equation}
		\mathbb{P}\left[M\left(x\right)\in S\right]\le e^{\epsilon}\mathbb{P}\left[M\left(x'\right)\in S\right]+\delta\text{ for all }S\in\Sigma_{2}^{q}.\tag*{$\triangle$}
		\end{equation}
\end{dfn}

At time~$k$, the system has state~$x(k)\in \mathbb{R}^{n}$, with discrete-time  dynamics 
\begin{align}\label{eq:dynamics}
x(k+1) &= H x(k)+w(k)\\
y(k) &= C x(k),
\end{align}
where process noise for the system is denoted by $w\left(k\right)\in\mathbb{R}^{n}$ and the matrices $H\in\mathbb{R}^{n\times n}$ and $ C\in \mathbb{R}^{q\times n} $ are  time-invariant. The probability distribution of the process noise is given by~$w\left(k\right)\sim\mathcal{N}\left(0,W\right)$, 
where $0\prec W\in\mathbb{R}^{n\times n}$, and all process noise terms are assumed to have finite variance. We assume that the matrices $ H $, $ W $, and $ C $ are public information, representing, e.g., that the device producing outputs is of a known type.

At each time $k$, the system outputs the value $y\left(k\right)$. 
Absent any privacy protections, the values of~$y$ could reveal those of~$x$ over time, which would compromise the system's privacy. Therefore, noise
must be added to the system's output to protect its state trajectory. That is, $ y $ is what is shared, $ x $ is what is sensitive, and because $ y $ is a function of $ x $, we add noise to $ y $ to protect $ x $. Calibrating the level of noise is done using the ``sensitivity'' of a system's output
map, which we define next for the input perturbation privacy we use; we emphasize that, although the system perturbs the outputs of its own dynamics, the ``input
perturbation'' label applies because the system perturbs what will become the inputs to a Kalman filter. 
The following bound is adapted from~\cite[Section IV-A]{LeNy2014}.

\begin{dfn}\label{dfn:sensitivity for dynamic systems}
	\emph{(Sensitivity for Input Perturbation Privacy)} 
		The $\ell_{2}\text{-norm}$ sensitivity of
		a system's output map is the greatest distance
		between two output trajectories which correspond to adjacent state
		trajectories. Formally, for $x,x'\in\tilde{\ell}_2^{n}$,
		\begin{equation}
		\Delta_{\ell_{2}}y: =\underset{x,x'\mid\textnormal{Adj}_{B}(x,x')=1}{\sup}\left\Vert Cx-Cx'\right\Vert _{\ell_{2}}. \tag*{$\triangle$}
		\end{equation}
\end{dfn}
We can bound $ \Delta_{\ell_{2}}y $ via $ \Delta_{\ell_{2}}y \le s_1 (C) B $ \cite{hale2018}, where $ s_1(\cdot) $ is the maximum singular value of a matrix.
Various mechanisms have been developed for enforcing differential privacy in the literature \cite{Dwork2014algorithmic}. The Gaussian mechanism requires adding  Gaussian noise to outputs to mask systems' state trajectories, and it can be useful in control settings that are robust to Gaussian noise. We next provide a definition of the Gaussian mechanism in terms of the $ \mathcal{Q} $-function, defined by $
\mathcal{Q}\left(y\right)=\frac{1}{\sqrt{2\pi}}\int_{y}^{\infty}e^{-\frac{z^{2}}{2}}dz
$.
\begin{lemma}\label{lem:gaussian mechanism}
\emph{(Input Perturbation Gaussian Mechanism)}
Let $\epsilon>0$ and $\delta\in\left(0,\nicefrac{1}{2}\right)$ be given. 
Let~$y \in \tilde{\ell}_2^{q}$ denote the output of a system with state trajectories
in~$\tilde{\ell}_{2}^{n}$,
and denote its ${\ell}_{2}$-norm sensitivity by $\Delta_{\ell_{2}}y$.
Then the Gaussian mechanism for $\left(\epsilon,\delta\right)$-differential privacy
takes the form
\begin{equation}
\tilde{y}(k)=y(k)+v(k),
\end{equation}
where~$v$ is a stochastic process with $v(k)\sim\mathcal{N}\left(0,\sigma^{2}I_{q}\right)$,
$I_{q}$ is the $q\times q$ identity matrix, and 
\begin{equation}
\sigma\ge\frac{\Delta_{\ell_{2}}y}{2\epsilon}\left(K_{\delta}+\sqrt{K_{\delta}^{2}+2\epsilon}\right)\text{ where }K_{\delta}:=\mathcal{Q}^{-1}\left(\delta\right). 
\end{equation}
This Gaussian mechanism provides $(\epsilon,\delta)$-differential privacy.
\end{lemma}

	\emph{Proof:}	See  \cite[Corollary 1]{LeNy2014}. \hfill $\blacksquare$
	
\noindent In words, the Gaussian mechanism adds i.i.d Gaussian
noise point-wise in time to the output of a system to keep its state trajectory private. We will use the Gaussian mechanism to enforce differential privacy for the remainder of the paper.

\subsection{Problem Formulation}
Having covered the relevant privacy background, we now state the problem that is the focus of the paper. 

\label{subsec:problem}
\begin{problem}\label{prb:problem1}
	Consider a system with publicly known mean initial condition~$\hat{x}^-(0)$, and let
    it have dynamics 
	\begin{align}
	x(k+1)&=Hx(k)+w(k)\\
	y(k)& = Cx(k).
	\end{align}
	Keep the state trajectories of the system differentially private according to specified privacy parameters $(\epsilon,\delta)$. 
	Next, investigate the effects of privacy in the following ways:
	
	(a) Given privacy parameters $ (\epsilon,\delta) $, quantify the ability of the recipients of the privatized outputs to accurately estimate the actual state trajectories of the system as a function of $ \epsilon$ and $ \delta $.
	
	(b) Develop guidelines for choosing the system's privacy parameters $ (\epsilon,\delta) $ to achieve pre-specified bounds on filter error. \hfill $\diamond$
\end{problem}
We will examine Problem 1(a) by quantifying filtering error and entropy in terms of systems' privacy parameters. Problem 1(b) will then use these
error bounds to inform how systems select their privacy parameters. 

The initial state of the system is denoted by $ \hat{x}^{-}(0) = \mathbb{E}[x(0)] $, where the minus sign will be used to initialize a Kalman filter which will be defined formally later. The next section presents our privacy implementation.

\section{Private Filtering Implementation}\label{sec:privacyImplementation}
We consider scenarios in which systems share their privatized outputs with a recipient, such as a utility company in a smart power grid or a traffic monitor in a smart transportation system. Abstracting away implementation details, we simply say that a system sends its outputs to an aggregator, and this aggregator will run a Kalman filter. 
To protect its own privacy, the system only shares its privatized outputs with the aggregator. The privatized outputs of the system may also be received by other entities e.g., other systems in a network, adversaries, an eavesdropper, and data analysts. Our results apply to these other recipients as well.

Without privacy, this transmission of data could reveal the system's state trajectory, and, as a result, compromise the system's privacy. 
Hence, the system adds privacy noise at each time $k$ to its output before sharing it, giving
\begin{equation}\label{eq:output}
\tilde{y}\left(k\right):=y\left(k\right)+v\left(k\right)=Cx\left(k\right)+v\left(k\right),
\end{equation}
where the privacy noise $v\left(k\right)\sim\mathcal{N}\left(0,\sigma^{2}I_{q}\right)$ as in Lemma~\ref{lem:gaussian mechanism}. 
 Introducing privacy naturally involves sacrificing a level of accuracy in the shared data, and the trade-offs and effects of privacy need to be rigorously evaluated to quantify the performance of private filtering.

The aggregator receives the privatized outputs of the system and implements a Kalman filter. The  Kalman filter minimizes the mean squared error (MSE) of both prediction and estimation for the systems studied in this paper, which
are linear systems with Gaussian noise. Mathematically, the Kalman filter minimizes both 
\begin{equation}
E\big[\Vert x(k)-\hat{x}^{-}(k)\Vert^{2}\big] \text{ and } E\big[\Vert x(k)-\hat{x}(k)\Vert^{2}\big],
\end{equation} 
where $ \hat{x}^{-}(k) $ and $ \hat{x}(k) $ respectively denote the \emph{a priori} state prediction and \emph{a posteriori} state estimate of the Kalman filter
at time $ k $. 
As noted in Problem~\ref{prb:problem1}, the term $\hat{x}^{-}(0)$ is assumed to be publicly known. We consider a steady-state Kalman filter because we are interested in systems over long time horizons. 

The update equation for the prediction step of the Kalman filter is evaluated as \cite{Welch1995} 
\begin{equation}\label{kalmanfilterprior} 
\hat{x}^{-}(k+1)=H\hat{x}(k),
\end{equation}
 and the \emph{a posteriori} state estimate $ \hat{x}(k) $ is updated as
\begin{equation}\label{eq:kalmanfilter}
\hat{x}(k+1)= \hat{x}^{-}(k+1)+\barsig C^{T}V^{-1}(\tilde{y}(k+1)-C \hat{x}^{-}(k+1)).
\end{equation}
Assuming the observability of the pair $ (H,C) $ and controllability of the pair $ (
A,D)  $, where $ W = DD^T $, the \emph{a posteriori} error covariance matrix $\barsig$ is computed as
\begin{align}\label{eq:sigmabar}
\barsig&=\Sigma-\Sigma C^{T}\left(C\Sigma C^{T}+V\right)^{-1}C\Sigma \nonumber \\ 
        &=(C^{T}V^{-1}C+\Sigma^{-1})^{-1}.
\end{align}
Here, the \emph{a priori} error covariance matrix $\Sigma$ is the unique positive semidefinite solution
to the discrete algebraic Riccati equation
\begin{align}\label{eq:DARE}
\Sigma&=H\Sigma H^{T}-H\Sigma C^{T}\left(C\Sigma C^{T}+V\right)^{-1}C\Sigma H^{T}+W \nonumber \\ 	
&=H\left(\Sigma^{-1}+ C^{T}V^{-1}C\right)^{-1}H^{T}+W.
\end{align}

Under differential privacy, it is provably unlikely for the recipients of  $ \tilde{y}(k) $ to distinguish a system's actual state trajectory from an adjacent one. In this setting, the Kalman filter minimizes the error in state prediction and estimation in the mean square sense, which means it provides the optimal estimate of a private state trajectory. Therefore, studying the connection between the Kalman filter and data privacy can elucidate fundamental limits of information accuracy when dealing with private trajectories.

\section{Quantifying the Effects of Privacy}\label{sec:results}
In this section, we explicitly quantify the ability of the aggregator or any potential recipient of private data, e.g., an adversary or an eavesdropper, to uncover the state trajectory of a system using its privatized outputs. One natural way to do so is to bound the MSE of the prediction and estimation steps of a Kalman filter that processes private data. Bounding these errors as functions of the system's privacy parameters will directly connect the system's privacy levels to the accuracy with which the aggregator can estimate its state values. 
We proceed by developing trace bounds for the \emph{a priori} error covariance matrix~$\Sigma$ and the \emph{a posteriori} error covariance matrix~$\overline{\Sigma}$, which are equal to the MSE of the prediction and MSE of the estimate in the Kalman filter, respectively. Because the Kalman filter in steady state minimizes both of these quantities, lower bounds on them are lower bounds on MSE for \emph{any} filtering strategy across long time horizons. 

Toward doing so, the following lemma upper and lower bounds the trace of a matrix product. 
In it, we use $ \lambda_{n}(K)\le \dots \le \lambda_{1}(K) $ to denote the eigenvalues of the matrix $ K $.

\begin{lemma}\label{lem:TraceofProduct}
	Let $K$ and $S$ be $n\times n$ matrices. If $K=K^{T}\geq0$
	and $S$ is symmetric, then 	
	\begin{equation}
	\lambda_{n}(S)\textnormal{tr}(K)\leq\textnormal{tr}(KS)\leq\lambda_{1}(S)\textnormal{tr}(K).
	\end{equation}
\end{lemma}

\begin{IEEEproof}
	See \cite[Fact 5.12.4]{Bernstein2009}. \hfill $\blacksquare$
\end{IEEEproof}

We next have an analogous lemma for matrix sums. 
\begin{lemma}\label{lem:eigenvalueSplit}
	Let $ K $ and $ S $ be $ n\times n $ Hermitian matrices. Then
	 \begin{align}\label{key}
	  \lambda_1(K)+\lambda_n (S)  &\le\lambda_{1}(K+S) \le \lambda_1 (K) + \lambda_1 (S)\\
	  	 \lambda_n(K)+\lambda_n (S)  &\le\lambda_{n}(K+S) \le \lambda_ n(K) + \lambda_1 (S).
	 \end{align}
\end{lemma}
\begin{IEEEproof}
See \cite[Theorem 8.4.11.]{Bernstein2009}. \hfill $\blacksquare$
\end{IEEEproof}

To ease the presentation of the forthcoming results, we impose the following assumption.
\begin{assumption}
	The matrix $C$ is diagonal.
\end{assumption}
Below, we will repeatedly encounter the term~$C^TV^{-1}C$, and we present bounds that  we will use below.
First observe that
\begin{equation}
C^{T}V^{-1}C=\text{diag}\left(\frac{C_{11}^{2}}{\sigma_{1}^{2}},\dots,\frac{C_{nn}^2}{\sigma_{n}^{2}}\right), 
\end{equation}
and we define
\begin{equation}
\begin{cases}
l:=\arg\min_{1\le i\le n}\frac{C_{ii}^{2}}{\sigma_{i}^{2}}\\
u:=\arg\max_{1\le i\le n}\frac{C_{ii}^{2}}{\sigma_{i}^{2}}
\end{cases}
\end{equation}
and then 
\begin{equation}\label{eq:CuSigmau}
\begin{cases}
\lambda_{n}\left(C^{T}V^{-1}C\right) & =\frac{C_{l}^{2}}{\sigma_{l}^{2}}\\
\lambda_{1}\left(C^{T}V^{-1}C\right) & =\frac{C_{u}^{2}}{\sigma_{u}^{2}}.
\end{cases}
\end{equation}
	
Next, we present lower and  upper bounds for the \emph{a priori} error covariance of the Kalman filter as functions of system's privacy noise. 

\begin{theorem}\label{thm:Sigmabounds}
	The steady state \emph{a priori} error covariance of the Kalman filter, equal to $ \textnormal{tr} \Sigma $, is bounded via
	\begin{equation} 
	 \textnormal{tr}W+\frac{\sigma_{u}^2\textnormal{tr}(H^{T}H)\lambda_{n}(W)}{\sigma_{u}^2+\lambda_{n}(W)C_u^2} \le \textnormal{tr}\Sigma \le \textnormal{tr}W+  \frac{\sigma_{l}^2\textnormal{tr}(H^{T}H)}{C_l^2}.
	\end{equation}
\end{theorem}

\begin{IEEEproof}
	The steady state MSE of the predictions of the Kalman filter is equal to the trace of the  \emph{a priori} error covariance of the Kalman filter as given in Equation~\eqref{eq:DARE}.
	Taking the trace of Equation~\eqref{eq:DARE}, we obtain
	\begin{align}
	\text{tr}\Sigma-\text{tr}W	&=\text{tr}\Big[H(\Sigma^{-1}+C^{T}V^{-1}C)^{-1}H^{T}\Big]
	\\&=\text{tr}\Big[H^{T}H(\Sigma^{-1}+C^{T}V^{-1}C)^{-1}\Big],
	\end{align}
	where we have used the cyclic permutation property of the trace. Next, we use Lemma~\ref{lem:TraceofProduct} to write
	\begin{align}
	\text{tr}\Sigma-\text{tr}W & \ge\text{tr}(H^{T}H)\lambda_{n}\Big[(\Sigma^{-1}+C^{T}V^{-1}C)^{-1}\Big]\\
	& =\frac{\text{tr}(H^{T}H)}{\lambda_{1}\left(\Sigma^{-1}+C^{T}V^{-1}C\right)}\\
	& \ge\frac{\text{tr}(H^{T}H)}{\lambda_{1}(\Sigma^{-1})+\lambda_{1}\big(C^{T}V^{-1}C\big)}\\
	& =\frac{\text{tr}(H^{T}H)}{\frac{1}{\lambda_{n}(\Sigma)}+\lambda_{1}\big(C^{T}V^{-1}C\big)},
	\end{align}
	where we apply Lemma~\ref{lem:eigenvalueSplit} on the third line to split up the eigenvalues and use the fact that $ \lambda_{1}(\Sigma^{-1})=\nicefrac{1}{\lambda_{n}(\Sigma)} $ in the final step. It is shown in \cite[Theorem 3.1]{Garloff86} that $\Sigma\ge W$, and therefore $\lambda_n(\Sigma)\ge \lambda_n(W)$. Using this fact and Equation~\eqref{eq:CuSigmau}, we find
	\vspace{-1 em}
	\begin{align}
	\text{tr}\Sigma-\text{tr}W\ge\frac{\sigma_{u}^2\textnormal{tr}(H^{T}H)\lambda_{n}(W)}{\sigma_{u}^2+\lambda_{n}(W)C_u^2},
	\end{align}
	which completes the first part of the proof. Similarly, by applying Lemmas~\ref{lem:TraceofProduct} and \ref{lem:eigenvalueSplit}  consecutively to Equation~\eqref{eq:DARE}, $ \text{tr}\Sigma $ can be upper-bounded as 
	\begin{align}
	\text{tr}\Sigma-\text{tr}W & \le\text{tr}(H^{T}H)\lambda_{1}\Big[(\Sigma^{-1}+C^{T}V^{-1}C)^{-1}\Big]\\
	& =\frac{\text{tr}(H^{T}H)}{\lambda_{n}\left(\Sigma^{-1}+C^{T}V^{-1}C\right)}\\
	& \le \frac{\text{tr}(H^{T}H)}{\lambda_{n}(\Sigma^{-1})+\lambda_{n}(C^{T}V^{-1}C)} \le \frac{\sigma_{l}^2\text{tr}(H^{T}H)}{C_l^2},
	\end{align}
	where in the second step we have used $ \lambda_{1}(M^{-1})=\nicefrac{1}{\lambda_{n}(M)} $  and the third step uses Lemma~\ref{lem:eigenvalueSplit} to split the eigenvalues. This completes the proof. \hfill $\blacksquare$
\end{IEEEproof}

Theorem~\ref{thm:Sigmabounds} bounds the MSE of the aggregator's prediction of a system, which quantifies the ability of the aggregator to infer future states of the system. The following theorem presents similar bounds for $ \overline{\Sigma} $, which represent the aggregator's ability to determine the system's
current state. 

\begin{theorem}\label{thm:traceboundSigmaBar}
	Suppose a system shares its privatized output trajectory, and the aggregator has all public information. Then, the steady-state MSE of the \emph{a posteriori} estimate of the system's states is bounded by
	\begin{equation}
	\frac{n\sigma_{u}^2}{C_u^2+\sigma_{u}^2\lambda_{n}^{-1}(W)} \le \text{tr}\overline{\Sigma}  \leq n\frac{\sigma_{l}^2}{C_l^2}. 
	\end{equation}
\end{theorem}
\begin{IEEEproof}
	 The steady-state mean-squared estimation error $ E\big[\Vert x(k)-\hat{x}(k)\Vert^{2}\big] $ is equivalent to the trace of the \emph{a posteriori} error covariance matrix $ \overline{\Sigma} $ in Equation~\eqref{eq:sigmabar}.  
	Using Lemma~\ref{lem:TraceofProduct}, a lower bound for the trace of  $\overline{\Sigma}$ can derived as
	\begin{align*}
	\text{tr}\overline{\Sigma} & \geq n \lambda_{n}\big((C^{T}V^{-1}C+\Sigma^{-1})^{-1}\big)  = \frac{n}{\lambda_{1}(C^{T}V^{-1}C+\Sigma^{-1})}\\
	& \geq\frac{n}{\lambda_{1}(C^{T}V^{-1}C)+\lambda_{1}(\Sigma^{-1})}\\
	& = \frac{n}{\lambda_{1}(C^{T}V^{-1}C)+\frac{1}{\lambda_{n}(\Sigma)}}\\
	& \geq \frac{n}{\lambda_{1}(C^{T}V^{-1}C)+\lambda_{n}^{-1}(W)}\\
	& =\frac{n\sigma_{u}^2}{C_u^2+\sigma_{u}^2\lambda_{n}^{-1}(W)},	
	\end{align*}
	where in the second line we have used Lemma~\ref{lem:eigenvalueSplit} to split the eigenvalues. In the second-to-last line,  we use $\Sigma\ge W$ based on \cite[Theorem 3.1]{Garloff86} to use $\lambda_n(\Sigma)\ge \lambda_n(W)$. 
	Similarly, using Lemma~\ref{lem:TraceofProduct}, an upper bound for the trace of $\overline{\Sigma}$ can be derived as
	\begin{align*}
	\text{tr}\overline{\Sigma} & \leq n \lambda_{1}\big((C^{T}V^{-1}C+\Sigma^{-1})^{-1}\big)\\
	& \leq\frac{n}{\lambda_{n}(C^{T}V^{-1}C+\Sigma^{-1})}\\
	& \leq\frac{n}{\lambda_{n}(C^{T}V^{-1}C)+\lambda_{n}(\Sigma^{-1})}\\
	& \leq\frac{n}{\lambda_{n}(C^{T}V^{-1}C)},
	\end{align*}
where in the last line $ \lambda_{n}(\Sigma^{-1}) >0 $ is eliminated. \hfill $\blacksquare$
\end{IEEEproof}

Together, the upper and lower bounds on $ \text{tr}(\overline{\Sigma})  $ give MSE bounds which elucidate the balance between privacy and accuracy of information shared with the aggregator.

Privacy and utility can be inherently conflicting goals, in the sense that the greater the level of privacy is, the less useful information will generally be. To study this relationship, we use an information theoretic tool to investigate the effects of the privacy noise $ v(k) $. In particular, we consider the differential entropy in the \emph{a posteriori} estimates $ \hat{x}(k) $ and \emph{a priori} predictions $ \hat{x}^{-}(k) $, which were defined  in Section~\ref{sec:privacyImplementation}. Shannon entropy has been used to investigate the leakage of information while using differential privacy in other settings, for example in \cite{alvim2012} and in distributed
linear control systems \cite{Wang2014}. Differential entropy  is useful for  Gaussian
distributions because it bounds the sub-level sets of
$\mathcal{R}^{-1}$, where $\mathcal{R}\left(y\right)=1-2\mathcal{Q}\left(y\right)$,
which is the volume of a covariance ellipsoid. Therefore, we will quantify
the effects of privacy noise upon the aggregator by studying
how privacy noise affects $\ln\det\Sigma$ and  $\ln\det\overline{\Sigma}$, which
are within an additive and multiplicative factor of the differential entropy of error in $\hat{x}$ and $ \hat{x}^- $, respectively.  Next, we present log-determinant bounds for the \emph{a priori} error covariance of the Kalman filter.

\begin{theorem}\label{thm:logdetSigma}
	Suppose that
	\begin{equation}
	s_{1}^{2}\left(H\right)<1+\left(s_{n}^{2}\left(H\right)\lambda_{1}\left(\Gamma\right)+\lambda_{n}\left(W\right)\right)\cdot\frac{C_{l}^{2}}{\sigma_{l}^2}
	\end{equation}
	where $\Gamma=\text{diag}\left(\gamma_{1},\dots,\gamma_{n}\right)$,
	$
	\gamma_{i}:=\frac{\sigma_{i}^{2}W_{ii}}{\sigma_{i}^{2}+C_{ii}^{2}W_{ii}},
	$ and ${ s_1(\cdot)\ge \dots \ge  s_n(\cdot)}$ denote the singular values of a matrix.
	The log-determinant of the \emph{a priori} error covariance of the Kalman filter can be upper-bounded as 
	\begin{equation} \ln\det\Sigma<\left(\frac{\sigma^2_{l}\lambda_{1}\left(W\right)}{\sigma^2_{l}+\eta  C^2_{l}-\sigma^2_{l}s_{1}^{2}\left(H\right)}\right)\sum_{i=1}^{n}s_{i}^{2}\left(H\right)+\textnormal{tr}W,
	\end{equation}
	where
	$
	\eta:= s_{n}^{2}\left(H\right)\lambda_{1}\left(\Gamma\right)+\lambda_{n}\left(W\right).
	$
	Furthermore, the log-determinant of the \emph{a priori} error covariance of the Kalman filter can be lower-bounded as
	\begin{equation}
	\ln\det\Sigma\ge\ln\Bigg[\frac{\sigma^2_{u}(\det H)^{2}}{\sigma^2_{u}\lambda^{-1}_n(W)+C^2_{u}+\sigma^2_{u}\ln n}+\det(W)\Bigg].
	\end{equation}
\end{theorem}
\begin{IEEEproof}
	See \cite{hale2018}.  \hfill $\blacksquare$
\end{IEEEproof}

Next, we derive bounds on the log-determinant of the \emph{a posteriori} error covariance of the Kalman filter. To facilitate the following analysis let us define the function
\begin{equation}
f\left(X\right)=X-XC^{T}(CXC^{T}+V)^{-1}C\label{eq:f(x)},
\end{equation}
where $X=X^T\succeq0$ is a variable and the matrices $ C  $ and $ V $ were defined in Section~\ref{sec:privacyReview_problemFormulation}. We state the following elementary lemmas that we will use below.
\begin{lemma}\label{lem:tA>B}
	If $A\succ0$ and $B\succeq0,$ then there exists $t\geq0$ such
	that $tA\geq B$.
\end{lemma}

\begin{IEEEproof}
	See \cite[Lemma 3]{Murray2005}. \hfill $\blacksquare$
\end{IEEEproof}

\begin{lemma}
	For every $t\geq0,$ we have 
	$f(tC^{-1}VC^{-T})\preceq C^{-1}VC^{-T}$.
\end{lemma}

\emph{Proof}:
	Inspired by the work in \cite{Murray2005}, let $t\ge0$. Then,
	\begin{equation}
	f(tC^{-1}VC^{-T}) =\frac{t}{t+1}C^{-1}VC^{-T} \preceq C^{-1}VC^{-T}. \tag*{ $\blacksquare$}
	\end{equation}
\begin{lemma}\label{lem:PDeigenvalue}
	Let $S$ be an $ n\times n $ Hermitian matrix. Then, 
	\begin{equation}
	\lambda_{\min}(S)I\preceq S\preceq\lambda_{\max}(S)I.
	\end{equation}
\end{lemma}

\begin{IEEEproof}
	See \cite[Corollary 8.4.2]{Bernstein2009}. \hfill $\blacksquare$
\end{IEEEproof}

We now present our log-determinant bounds for $ \overline{\Sigma} $.

\begin{theorem}\label{thm:logdetSigmaBar}
	The log-determinant of the \emph{a posteriori} error covariance of the Kalman filter is bounded via 
	\begin{equation}
	n \ln\left(  \frac{\sigma_{u}^2}{C_u^2+\sigma_{u}^2\lambda_{n}^{-1}(W)}\right) \leq \ln\det \overline{\Sigma}\leq n \ln \left( \frac{\sigma_l^2}{C_l^2}\right) .
	\end{equation}
\end{theorem}
\begin{IEEEproof} Computing~$f(\Sigma)$ in Equation~\eqref{eq:f(x)}, we get 
\begin{equation}
\overline{\Sigma}=f(\Sigma)=\Sigma-\Sigma C^{T}(C\Sigma C^{T}+V)^{-1}C\Sigma.
\end{equation}
 By Lemma~\ref{lem:tA>B}, there
exists a $t\geq0$ such that~$\Sigma\preceq tC^{-1}VC^{-T}$
because $ \Sigma \succeq 0 $ by definition and $ C^{-1}VC^{-T} \succ 0$. Since $f$ is a monotonic function \cite{Murray2005}, we 
have~$f(\Sigma)\preceq f(tC^{-1}VC^{-T})$,
and therefore by Lemma 2,
\begin{align}
\overline{\Sigma} =f(\Sigma)\preceq f(tC^{-1}VC^{-T})\preceq C^{-1}VC^{-T}.
\end{align}	 
Taking the log-determinant of both sides, we find 
\begin{align}\label{key}
\ln\det \overline{\Sigma}\leq \ln\det \left(C^{-1}VC^{-T} \right)  = \ln \prod_{i=1}^{n} \frac{\sigma^2}{C_{ii}^2}\le n \ln \left( \frac{\sigma_l^2}{C_l^2}\right) .
\end{align}
Next, using Lemma~\ref{lem:PDeigenvalue}, Equation~\eqref{eq:sigmabar} implies that
\begin{align}\label{eq:sigmabarPre}
&\overline{\Sigma}\succeq\lambda_{n}\left( (C^{T}V^{-1}C+\Sigma^{-1})^{-1}\right) I \nonumber\\ 
& \!\!\!= \!\! \frac{1}{\lambda_{1}(C^{T}V^{-1}C+\Sigma^{-1})}I \!\succeq\!  \frac{1}{\lambda_{1}(C^{T}V^{-1}C)+\lambda_{1}(\Sigma^{-1})}I \nonumber \\
& \!\!\!\! \!\!\!\succeq\!\!  \frac{1}{\lambda_{1}(C^{T}V^{-1}C)\!+\!\frac{1}{\lambda_{n}(\Sigma)}}I \!\succeq\!  
\frac{1}{\lambda_{1}(C^{T}V^{-1}C)\!+\!\lambda_{n}^{-1}(W)}I, 
\end{align}
where, due to the similarity of the steps of this proof to the proof for Theorem~\ref{thm:traceboundSigmaBar}, we have omitted the explanations for each step. Using Equation~\eqref{eq:CuSigmau} and taking the log-determinant of the both sides of the Equation~\eqref{eq:sigmabarPre}, we can write 
\begin{equation}
\ln \det \overline{\Sigma} \ge \ln \left[  \left(\frac{\sigma_{u}^2}{C_u^2+\sigma_{u}^2\lambda_{n}^{-1}(W)}\right)^n \right] ,
\end{equation}
\noindent and the theorem follows. \hfill $\blacksquare$
\end{IEEEproof} 

Of course, beyond merely studying the impacts of privacy, one can leverage these bounds to enable better privacy parameter selection by tailoring privacy levels to attain a certain quality of information downstream. That is the subject of the next section.

\section{Guidelines for Selecting Privacy Parameters}\label{sec:guidelines}

In this section, we develop new guidelines for selecting  privacy parameters, which will allow us to achieve specified filtering error bounds. These bounds enable the calibration of privacy levels based on the desired accuracy of those making decisions with private data, as well as individuals' privacy desires. The value of the privacy parameter $ \delta $ is typically chosen to be small and fixed. The value of $ \delta $ can be understood as the probability of differential privacy failing to protect sensitive data, and it is therefore often \cite{LeNy2014} chosen in $ [10^{-5},0.1] $ and we adopt this for range for the rest of the paper.

\begin{theorem}\label{thm:guidelineSigma}
	Suppose a system shares its privatized output trajectory and the aggregator has access to all  public information.  Take ${ \delta \in [10^{-5}, 10^{-1}] }$ and choose $ 	\sigma =\frac{\Delta_{\ell_{2}}y}{2\epsilon}\left(K_{\delta}+\sqrt{K_{\delta}^{2}+2\epsilon}\right) $. Suppose we want the steady state MSE of predictions of the system's next states, i.e., the \emph{a priori} state predictions, to be in $ [B_l,B_u] $ for some bounds $ B_l $ and $ B_u $. A sufficient condition to do so is to bound $ \epsilon $ via
	\begin{equation}\label{eq:epsilonGuidelines}
	\frac{1}{8}\left(\frac{1+\sqrt{36\eta_{3}+1}}{\eta_{3}}\right)^{2}\le \epsilon \le \frac{1}{\eta_{1}},
	\end{equation}
	where \begin{equation}\label{eq:eta1}
	\eta_{1}:=\left(\frac{\left(B_{l}-\textnormal{tr}W\right)\lambda_{n}\left(W\right)C_{u}^{2}}{\left(\Delta_{\ell_{2}}y\right)^{2}\left(\textnormal{tr}(H^{T}H)\lambda_{n}\left(W\right)-B_{l}+\textnormal{tr}W\right)}\right)^{\nicefrac{1}{2}}
	\end{equation}
	and 
	\begin{equation}
	\eta_{3}:=\left(\frac{(B_{u}-\text{tr}W)C_{l}^{2}}{\left(\Delta_{\ell_{2}}y\right)^{2}\textnormal{tr}(H^{T}H)}\right)^{1/2}.
	\end{equation}
	
\end{theorem}

\begin{IEEEproof}
	First, choose $ \epsilon\ge\frac{1}{8}\left(\frac{1+\sqrt{36\eta_{3}+1}}{\eta_{3}}\right)^{2} $ and solve for $ \eta_{3} $ to get~$\frac{9+\sqrt{2\epsilon}}{2\epsilon}\le\eta_{3}$.
	Taking $ \delta \in [10^{-5}, 10^{-1}] $ implies ${ K_{\delta} \in [1,4.5] }$. As a result, we can write $ \nicefrac{(2K_{\delta}+\sqrt{2\epsilon})}{2\epsilon}\le\nicefrac{(9+\sqrt{2\epsilon})}{2\epsilon} $ to get
	\begin{equation}\label{key}
	\frac{2K_{\delta}+\sqrt{2\epsilon}}{2\epsilon}\le\eta_{3}.
	\end{equation}
	Using the fact that $ \sqrt {K+S} \le \sqrt{K} + \sqrt{S} $, substituting for $ \eta_{3} $, squaring both sides and rearranging, we have 
	\begin{align}
	\left(\Delta_{\ell_{2}}y\right)^{2}\left(\frac{K_{\delta}+\sqrt{K_{\delta}^{2}+2\epsilon}}{2\epsilon}\right)^{2}\le\frac{(B_{u}-\text{tr}W)C_{l}^{2}}{\textnormal{tr}(H^{T}H)},
	\end{align}
	which implies that~$\sigma_{l}^{2}\le\frac{(B_{u}-\text{tr}W)C_{l}^{2}}{\textnormal{tr}(H^{T}H)}$.
	It then follows that
	\begin{equation}
	\text{tr}W+\frac{\sigma_{l}^{2}\textnormal{tr}(H^{T}H)}{C_{l}^{2}}\le B_{u},
	\end{equation}
	which, by comparing to Theorem~\ref{thm:Sigmabounds}, implies that $ \tr \Sigma \le B_u $.
	
	Next, choose $ \epsilon \le \nicefrac{1}{\eta_{1}} $. Given $  K_{\delta} \ge 1  $, we can write ${ \epsilon \le \nicefrac{K_{\delta}}{\eta_1} }$, and, rearranging the terms, we find 
	$
	\eta_{1}\le\frac{K_{\delta}}{\epsilon}.
	$
	Substitute for $ \eta_{1} $, square, and rearrange to get 
	\begin{equation}\label{key}
	\frac{\left(B_{l}-\textnormal{tr}W\right)\lambda_{n}\left(W\right)C_{u}^{2}}{\textnormal{tr}(H^{T}H)\lambda_{n}\left(W\right)-B_{l}+\textnormal{tr}W}\le \left(\Delta_{\ell_{2}}y\right)^{2}\left(\frac{K_{\delta}}{\epsilon}\right)^{2}.
	\end{equation}
	Now, $  \frac{K_{\delta}}{\epsilon} \le\frac{K_{\delta}+\sqrt{K_{\delta}^{2}+2\epsilon}}{2\epsilon} $ and therefore
	\resizebox{.5 \textwidth}{!} 
	{
	$\frac{\left(B_{l}-\textnormal{tr}W\right)\lambda_{n}\left(W\right)C_{u}^{2}}{\textnormal{tr}(H^{T}H)\lambda_{n}\left(W\right)-B_{l}+\textnormal{tr}W}\le 
	\left(\Delta_{\ell_{2}}y\right)^{2}\left(\frac{K_{\delta}+\sqrt{K_{\delta}^{2}+2\epsilon}}{2\epsilon}\right)^{2}$,
	}
	which implies
	\begin{equation}\label{key}
	\frac{\left(B_{l}-\textnormal{tr}W\right)\lambda_{n}\left(W\right)C_{u}^{2}}{\textnormal{tr}(H^{T}H)\lambda_{n}\left(W\right)-B_{l}+\textnormal{tr}W}\le\sigma_{u}^{2}.
	\end{equation}
	Therefore, 
	\begin{equation}\label{key}
	\textnormal{tr}W+\frac{\sigma_{u}^2\textnormal{tr}(H^{T}H)\lambda_{n}(W)}{\sigma_{u}^2+\lambda_{n}(W)C_u^2} \ge B_l,
	\end{equation}
	and by Theorem~\ref{thm:Sigmabounds}, choosing $ \epsilon $ as above is sufficient to guarantee $ \tr \Sigma \ge B_l $. \hfill $\blacksquare$
\end{IEEEproof}

Theorem~\ref{thm:guidelineSigma} presents upper and lower bounds for the privacy parameter $ \epsilon$ which ensure that steady-state \emph{a priori} filtering error remains within acceptable bounds. Next, we provide analogous bounds on $ \epsilon $ for \emph{a posteriori} error. 

\begin{theorem}\label{thm:guidelineSigmaBar}
		Suppose a system shares its privatized output trajectory and the aggregator has access to all  public information. Take $ \delta \in [10^{-5}, 10^{-1}] $ and set ${ \sigma =\frac{\Delta_{\ell_{2}}y}{2\epsilon}\left(K_{\delta}+\sqrt{K_{\delta}^{2}+2\epsilon}\right) }$. Suppose we want the steady-state MSE of the estimated states of the system, i.e., the \emph{a posteriori} state estimates, to be contained in the interval $ [B_l,B_u] $ for some bounds $ B_l $ and $ B_u $. To do so, it is sufficient to choose the privacy parameter $ \epsilon $ according to
		\begin{equation}
	    \frac{1}{8}\left(\frac{1+\sqrt{36\eta_{4}+1}}{\eta_{4}}\right)^{2}\le \epsilon \le  \frac{1}{\eta_{2}},
		\end{equation}
		where 
		\begin{equation}\label{eq:eta2}
		\eta_{2}\!:=\!\left(\!\frac{B_{l}C_{u}^{2}}{\left(\Delta_{\ell_{2}}y\right)^{2}\!\left(n\!-\!B_{l}\lambda_{n}^{-1}(W)\right)}\!\right)^{1/2}\!\!\!\!\!\!, \,
		\eta_{4}\!:=\! \left(\!\frac{B_{u}C_{l}^{2}}{n\left(\Delta_{\ell_{2}}y\!\right)^{2}}  \right)^{1/2}.
		\end{equation}
\end{theorem}

\begin{IEEEproof}
	Choose $ \epsilon \ge 	    \frac{1}{8}\left(\frac{1+\sqrt{36\eta_{4}+1}}{\eta_{4}}\right)^{2} $ and solve for $ \eta_{4} $  to get~$\frac{9+\sqrt{2\epsilon}}{2\epsilon}\le\eta_{4}$.
	Choosing $ \delta \in [10^{-5}, 10^{-1}] $ gives ${ K_{\delta} \in [1,4.5] }$. As a result, similar to the proof of Theorem~\ref{thm:guidelineSigma}, we can write $ \nicefrac{(2K_{\delta}+\sqrt{2\epsilon})}{2\epsilon}\le\nicefrac{(9+\sqrt{2\epsilon})}{2\epsilon} $ to get
	\begin{equation}\label{key}
	\frac{2K_{\delta}+\sqrt{2\epsilon}}{2\epsilon}\le \eta_{4}.
	\end{equation}
	Because $ \sqrt{K+S} \le \sqrt{K} + \sqrt{S} $, we can lower-bound the left-hand-side to write 
	\begin{equation}\label{key}
	\frac{K_{\delta}+\sqrt{K_{\delta}^{2}+2\epsilon}}{2\epsilon} \le \eta_{4}.
	\end{equation}
	Squaring, substituting in $ \eta_{4} $, and rearranging we get 
	\begin{equation}\label{key}
	\left(\Delta_{\ell_{2}}y\right)^{2}\left(\frac{K_{\delta}+\sqrt{K_{\delta}^{2}+2\epsilon}}{2\epsilon}\right)^{2}\le\frac{B_{u}C_{l}^{2}}{n},
	\end{equation}
	which is equivalent to
	$
	\sigma_{l}^{2}\le\frac{B_{u}C_{l}^{2}}{n},
	$
	which implies
	$
	\frac{n\sigma_l^2}{C_l^2}\le B_u.
	$
	Comparing this result to Theorem~\ref{thm:traceboundSigmaBar}, we  see that this is sufficient for  $ \textnormal{tr}\overline{\Sigma}\le B_{u} $.

	Next, choose $ \epsilon\le  \frac{1}{\eta_{2}} $. Given $ K_{\delta}\in [1,4.5] $, we may write
	$
	\eta_{2}\le\frac{K_{\delta}}{\epsilon}.
	$
	We substitute for $ \eta_{2} $ and square both sides to write 
	\begin{equation}\label{key}
    \frac{B_{l}C_{u}^{2}}{\left(\Delta_{\ell_{2}}y\right)^{2}\left(n-B_{l}\lambda_{n}^{-1}(W)\right)}\le\left(\Delta_{\ell_{2}}y\right)^{2}\left(\frac{K_{\delta}}{\epsilon}\right)^{2},
	\end{equation}
	and therefore, by upper-bounding the right-hand-side and rearranging we write 
	\begin{equation}\label{key}
    \frac{B_{l}C_{u}^{2}}{n-B_{l}\lambda_{n}^{-1}(W)}\le\left(\Delta_{\ell_{2}}y\right)^{2}\left(\frac{K_{\delta}+\sqrt{K_{\delta}^{2}+2\epsilon}}{2\epsilon}\right)^{2}.
	\end{equation}
	This in turn implies
	$
	\frac{B_{l}C_{u}^{2}}{n-B_{l}\lambda_{n}^{-1}(W)}\le\sigma_{u}^{2}.
	$
	Isolating $ B_l $ gives
	$
	\frac{\sigma_{u}^{2}n}{C_{u}^{2}+\sigma_{u}^{2}\lambda_{n}^{-1}(W)}\ge B_{l},
	$ 
	which, in light of Theorem~\ref{thm:traceboundSigmaBar}, is a sufficient condition to get $ \tr \overline{\Sigma} \ge B_l $. \hfill $\blacksquare$
\end{IEEEproof}

Theorem~\ref{thm:guidelineSigmaBar} provides guidelines for choosing the privacy parameters $ (\epsilon, \delta) $, which allows a user to make informed decisions for its level of privacy. We next demonstrate these bounds in practice.
\section{Case Study}\label{sec:casestudy}
In this section, we simulate a system with state $ x(k)\in \mathbb{R}^2 $ for all $ k $ and dynamics
\begin{equation}\label{key}
H = \left[\begin{array}{cc}
1 & 1\\
0 & 1
\end{array}\right] \textnormal{ , }C = \left[\begin{array}{cc}
1 & 0\\
0 & 1
\end{array}\right] \textnormal{, and } W = 10  I_{2\times 2}.
\end{equation}
We proceed to enforce input perturbation differential privacy as discussed in Section~\ref{sec:privacyImplementation}. We choose ${ (\epsilon,\delta) = (\ln3, 0.001) }$, which gives $ \sigma = 2.96 $. The privacy noise $v\left(k\right)\sim\mathcal{N}\left(0,\sigma^{2}I_{2 \times 2}\right)$ is added to the outputs $ y (k)$ at each time $ k $. The aggregator receives the private outputs pointwise in time and runs a Kalman filter, and we simulate this setup for $ 100 $ timesteps. 

\begin{figure}
	\begin{center}
		\includegraphics[width=.75\columnwidth]{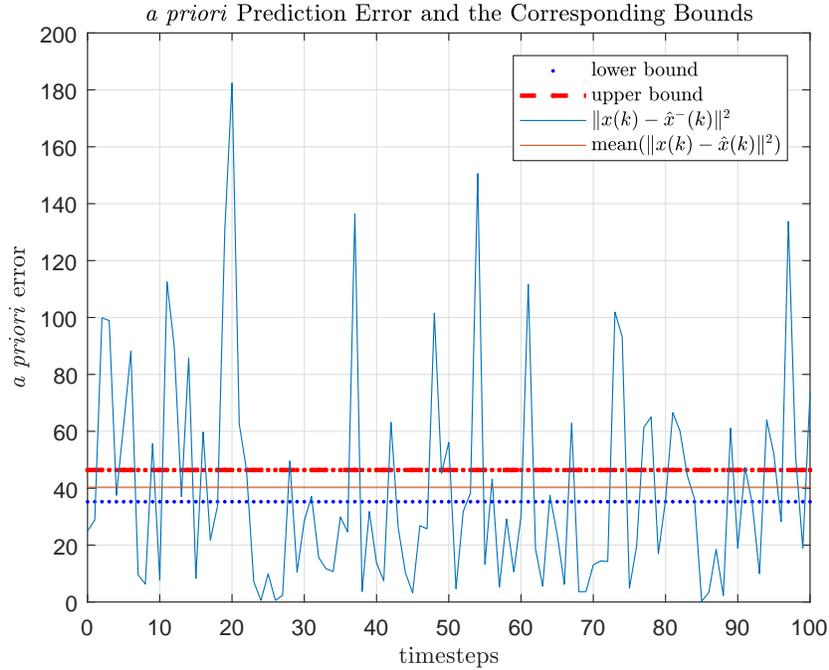}
	\end{center}
\vspace{-1.5 em}
	\caption{The squared error of the prediction of the aggregator (solid line), the lower bound on prediction error as developed in Theorem~\ref{thm:Sigmabounds} (dotted line), the upper bound on prediction error in Theorem~\ref{thm:Sigmabounds} (dashed line) over $100$ timesteps. Although our bounds are developed for mean-squared error, they hold at most timesteps for instantaneous error, and it is shown that on average the MSE lies within the bounds.  
	}
	\label{fig:priori_error}
\end{figure}

\begin{figure}
	\begin{center}
		\includegraphics[width=.75\columnwidth]{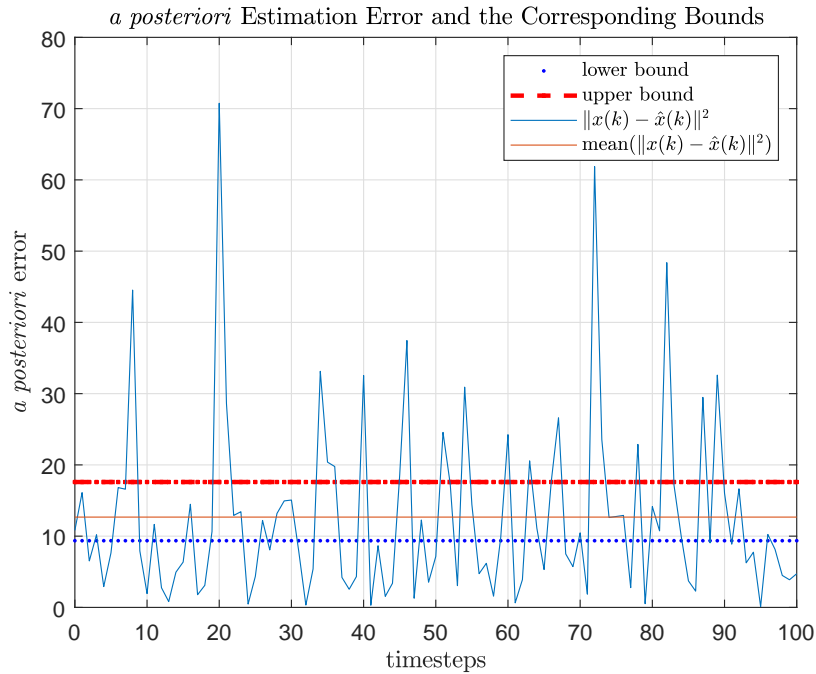}
	\end{center}
	\vspace{-1.5 em}
	\caption{The squared error of the estimation of the aggregator on system's states (solid line), the lower bound on estimation error as developed in Theorem~\ref{thm:traceboundSigmaBar} (dotted line), the upper bound on estimation error in Theorem~\ref{thm:traceboundSigmaBar} (dashed line) over $100$ timesteps. As in Figure~\ref{fig:priori_error}, we see that instantaneous \emph{a posteriori} error also typically obeys our MSE bounds, and on average lies within the bounds, as shown here.
	}
	\label{fig:post_error}
\end{figure}

The results of this simulation are presented in Figures \ref{fig:priori_error} and~\ref{fig:post_error}. In Figure \ref{fig:priori_error}, we present the MSE bounds derived in Theorem~\ref{thm:Sigmabounds}, and we compare them with the actual instantaneous \emph{a priori} error. On average, the \emph{a priori} error in predictions of the system's states remains within the given bounds; ephemeral bound violations are expected as these bounds pertain to mean-square error. In Figure \ref{fig:post_error}, we demonstrate the instantaneous error of the estimated states of the system and we compare that with the upper and lower bounds derived in Theorem~\ref{thm:traceboundSigmaBar}.   As expected, the instantaneous \emph{a posteriori} error typically lies within the bounds derived in Theorem~\ref{thm:traceboundSigmaBar}. Both plots illustrate our bounds on the ability of an aggregator to predict or estimate the states of a system sharing privatized information. 

\section{Conclusions} \label{sec:conclusions}
In this paper, we have proposed new guidelines for calibrating the levels of privacy when enforcing differential privacy in linear systems with Gaussian noise. These guidelines were chosen to attain desired filtering error bounds, and novel bounds were presented for both filter entropy and filter error in terms of a system's privacy levels. Future work includes investigating general filtering techniques in which nonlinear systems are considered, with potential applications in smart power grids and autonomous systems.


\bibliographystyle{IEEEtran}{}
\bibliography{sources}


\end{document}